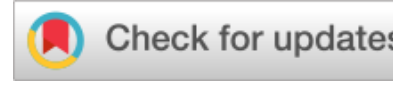

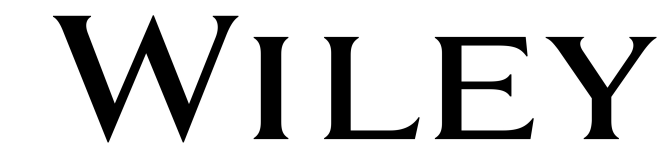

*Research Article*

# Symmetries in the Hamiltonian Formulation of String Theory

**Héctor A. Benítez** [1] **and René Negrón** [2]

[1] *Department of Engineering, Universidad Tecnológica del Perú, Lima, Peru*
[2] *Department of Engineering, Universidad de Lima, Lima, Peru*

Correspondence should be addressed to Héctor A. Benítez; hbenitez@utp.edu.pe

In the context of the Hamiltonian formulation of string theory, a widely acknowledged issue is the inability of the first-class constraints to accurately reproduce the Lagrangian symmetry transformations. We take a critical look at the Hamiltonian formulation of the Polyakov string and demonstrate, using Castellani's procedure, that diffeomorphism and Weyl symmetries naturally emerge without using any field-dependent reparametrization, thus preserving the theoretical consistency between Hamiltonian and Lagrangian descriptions. Additionally, we will review the standard Hamiltonian analysis of the symmetries in terms of lapse and shift functions and show that the reason behind this failure is not due to the noncanonicity of the variables but rather because this parametrization fails to preserve general covariance.

## 1. Introduction

In the early days of string theory, the presence of unphysical negative norm states posed a significant challenge. Virasoro was the first to identify an infinite set of conditions, the famous Virasoro algebra [1], which could effectively eliminate these unphysical states when imposed by an infinite set of operators. The origin of such a profound symmetry became clearer when Nambu [2] and later Goto [3] proposed an action describing the dynamics of the string. This action is proportional to the area of the so-called worldsheet spanned by the string on the target space. The arbitrariness in defining the local parameters characterizing the string dynamics results in an invariance under reparameterizations on the worldsheet. In the Nambu–Goto formulation, the Virasoro operators naturally emerge as the Fourier modes of the energy–momentum tensor. Early on, Goddard et al. [4] concluded that the spectrum is both unitary and Lorentz invariant at the critical dimension $d = 26$ in the light-cone and covariant gauge by applying the Dirac procedure for quantizing the Nambu–Goto string.

A few years later, Polyakov [5] introduced a quadratic and more flexible action for the string by promoting the worldsheet metric to a Lagrange multiplier. This action reproduces the classical equations of motion of the relativistic string and exhibits explicit invariance under diffeomorphisms. As a side effect, due to the arbitrariness in defining the worldsheet metric, the action possesses an additional symmetry, known as Weyl invariance (for a novel explanation of how the symmetries of the Polyakov string are derived using the Lagrangian formalism, we refer the reader to [6]).

When implementing the Hamiltonian formalism, the gauge invariance of the string action leads to a singular theory in which not all conjugate momenta are independent functions of the velocities. Dirac's pioneering work on constrained systems [7] laid the foundations for later systematic developments by Anderson and Bergman [8]. Building on this, Castellani [9] applied these techniques to field theories with varying lengths of constraint chains. Subsequent refinements have been made by Pons [10] and Rothe et al. [11].

For the Polyakov string, the Dirac analysis has been extensively reviewed in many articles and books (for instance, [12–14] and references therein). Two main features characterize the Hamiltonian formulation of the Polyakov string:

1. The absence of a kinetic term for the worldsheet metric leads to the vanishing of its conjugate momentum $\pi^{\alpha\beta}$, thereby serving as the primary constraints of the model.
2. The consistency condition for the time evolution of the primary constraints, $\dot{\pi}^{\alpha\beta} = 0$, does not result in three but only two independent secondary constraints. This is because the trace of $\dot{\pi}^{\alpha\beta}$ is identically zero.

Following Castellani's procedure, diffeomorphism and Weyl symmetries may emerge naturally in the Hamiltonian formulation of the relativistic string. However, this requirement has received surprisingly little attention, despite its importance in establishing the equivalence between Lagrangian and Hamiltonian formulations. Standard textbooks and lectures on string theory rarely address it explicitly. Three notable exceptions addressing this issue are those by Henneaux and Brink [14], Kuchař and Torre [15], and Pons et al. [16]. In the first case, the authors developed a comprehensive Hamiltonian analysis employing a specific parametrization of the worldsheet metric, introducing lapse and shift variables in an ADM (Arnowitt–Deser–Misner) fashion. Once the metric arbitrariness was removed, a gauge generator is proposed as a linear combination of the Virasoro constraints. Consistently, Kuchař and Torre showed that when the canonical variables are chosen as lapse and shift, the algebra of symmetry generators does not reproduce the worldsheet diffeomorphism algebra in a straightforward manner. Instead, it closes only after a field-dependent redefinition of the gauge parameters:

$$\nu^{\perp} = \varepsilon^0 N, \quad \nu^1 = \varepsilon^1 + N_1 \varepsilon^0, \tag{1}$$

which obscures the manifest covariance of the theory. In contrast, Pons et al. [16] developed a related method in which the primary constraints (the conjugate momenta of the metric) are modified so that the resulting secondary constraints coincide with the $T$ and $H$ Virasoro conditions once consistency is imposed. Using Castellani's procedure, the Hamiltonian generator of symmetries is then constructed from this set of constraints. Once more, the resulting transformations of both metric and matter fields do not match the form of diffeomorphism transformations, requiring instead the same field-dependent redefinition of the parameters as in (1).

It is noteworthy to mention that a similar characteristic emerged in the Hamiltonian formulation of general relativity, where diffeomorphism invariance is manifested in the Lagrangian framework. For many years, it was widely accepted that ADM decomposition was essential for studying the Hamiltonian formulation of general relativity. However, an awful result of this formulation is that diffeomorphism invariance is lost and just recovered by a field-dependent redefinition of gauge parameters [17, 18], leading to a clear contradiction of the equivalence between Hamiltonian and Lagrangian formulations [19].

The work of Samanta emphasized this disparity [20]. Building on Rothe's method for recovering gauge symmetries [21], the author derived the diffeomorphism transformations for the Einstein–Hilbert Lagrangian. Driven by this result, Kiriushcheva et al. [22] demonstrated that diffeomorphism invariance persists at the Hamiltonian level by applying the standard Castellani's procedure without any modification of the original Hamiltonian of GR. Moreover, they showed that the metric coordinates and ADM variables are related by a phase space transformation, which is not canonical [23]. Consequently, the error in finding the gauge symmetries for the Hamiltonian formulation of GR comes from using ADM variables in the Hamiltonian formulation [24].

This recurrent need for field-dependent redefinitions highlights a structural limitation of approaches based on ADM-type parametrizations. In what follows, we revisit the Hamiltonian analysis of bosonic string theory but strictly within a covariant chain of constraints, which allows Castellani's procedure to yield the correct diffeomorphism and Weyl symmetries without resorting to such redefinitions. To make this point transparent, we begin with the simpler case of the relativistic point particle. There we establish that reparametrization invariance is not the only possible local symmetry but rather a particular instance arising when the gauge parameters are chosen as worldline vectors. From this example, we learn two lessons: first, the failure to recover covariant transformations from first principles is not due to the use of noncanonical variables; second, Castellani's procedure requires the chain of constraints to be composed of covariant constraints. We then extend this reasoning to the relativistic string, where diffeomorphism and Weyl symmetries emerge naturally in the Hamiltonian formulation without imposing any specific parametrization of phase space variables. This result is essential for establishing the equivalence between Lagrangian and Hamiltonian formulations of the relativistic string. Finally, have revisited the Hamiltonian analysis in the lapse and shit variables. We discuss and prove that, regardless of the canonicity of these transformations, the generator of symmetries does not reproduce diffeomorphisms, and a field-dependent redefinition of the gauge parameters (1) is still required to recover the usual form of diffeomorphism transformations. Once again, this unexpected outcome is not a consequence of noncanonicity, but of choosing first-class constraints in Castellani's procedure that do not transform covariantly.

## 2. General Procedure

Let us consider a dynamical system described by a Lagrangian $L(q^i, \dot{q}^i)$. In the Lagrangian formulation, the classical dynamics can be obtained by studying the action

$$S = \int_{t_1}^{t_2} L(q, \dot{q}, t) dt. \tag{2}$$

The vanishing of the Euler derivatives leaves the action stationary

$$\mathscr{L}_i := \frac{\partial L}{\partial q^i} - \frac{d}{dt}\left(\frac{\partial L}{\partial \dot{q}^i}\right) = 0. \tag{3}$$

A gauge symmetry of the action is a transformation $\delta q^i = \varepsilon^r f^i_r(q, \dot{q})$, involving arbitrary functions of time $\varepsilon(t)$ that leave the action invariant, that is, $S[q(t)] = S[q(t) + \delta q(t)]$. In this way, if $\delta q(\varepsilon)$ is a symmetry of the theory, then the corresponding variation of the action should take the following form:

$$\delta S = -\int dt\, \varepsilon^r(t) \Lambda_r(q), \tag{4}$$

where $\Lambda$ is identically zero without the need of using the equations of motion.

When dealing with a Lagrangian $L$, it is crucial to investigate whether it exhibits a gauge symmetry. If it does, the equations of motion of the theory cannot be independent but rather related by a Noether identity. A technique developed by Shirzad [25] and Rothe et al. [21] exploits these relations to identify the gauge symmetries of the action.

Following Rothe, we start by examining the variation in the action when considering an infinitesimal arbitrary variation $\delta q^i$, that is,

$$\delta S = \int dt \mathscr{L}_i \delta q^i. \tag{5}$$

Now, let us consider the following transformations:

$$\delta q^i = \sum_{s=0}^{n} (-)^s \frac{d^s \varepsilon}{dt^s} \rho_{s,i}. \tag{6}$$

Substituting these transformations in (5), we find that the Noether identities $\Lambda$ that appear in (4) are given by

$$\Lambda = \sum_{s=0}^{n} \frac{d^s}{dt^s} \left( \rho_{s,i} \mathscr{L}_i \right). \tag{7}$$

Once we identify the $\rho$ functions that lead to the vanishing of the Noether identities in (7), we can derive the gauge transformations using (6).

In the Hamiltonian formalism, multiple methods are available for constructing the gauge generator once the canonical Hamiltonian and all the first-class constraints have been identified (reference [26] provides an instructive review of this subject). Given the canonical Hamiltonian $H_c$, the set of primary first-class constraints $\Phi^a$ and the associated Lagrange multipliers $\lambda$s, the total Hamiltonian $H_T$, defined by

$$H_T = H_c + \lambda^a \Phi_a, \tag{8}$$

describes the dynamics of the theory, leading to the time evolution vector field:

$$X_H = \frac{\partial}{\partial t} + \{H_T, \cdot\}. \tag{9}$$

So, the dynamical trajectories must satisfy the Hamiltonian equations of motion:

$$\dot{q}^i(t) = \left\{ q^i(t), H_T \right\}. \tag{10}$$

The ambiguity in defining the Lagrange multipliers $\lambda^a$ in (8) results in a family of trajectories, which are also classical solutions. Any two close trajectories must be related by an infinitesimal transformation as follows:

$$\delta q^i(t) = \left\{ q^i, G(t) \right\}, \tag{11}$$

where $G(t)$ is the generator of gauge transformations with explicit time dependence. According to Dirac [7], the generator of gauge transformations can be written as

$$G(t) = \sum_r \varepsilon^r(t) \Phi_r(q, p), \tag{12}$$

where the gauge parameters $\varepsilon^r$ are functions of time. In (12), the sum index runs over all the first-class constraints (not just the pfcc).

Nevertheless, not every generator of the form (12) produces symmetries of the Hamiltonian equations of motion. As shown by Castellani [9], $G(t)$ acts as a symmetry generator if and only if it corresponds to a conserved charge (see [17] for a pedagogical review of the subject). Thus,

$$\frac{\partial G}{\partial t} + \{G, H_T\} = \text{pfcc}. \tag{13}$$

This condition serves as the guiding principle for deriving the correct gauge transformations, thereby restricting the gauge parameters specified in (12) [21].

## 3. The Relativistic Point Particle

The relativistic particle traces a worldline parametrized by $\tau$, which maps on a target Minkowski space with coordinates $x^\mu(\tau)$. The action, which is given by

$$S = \frac{1}{2} \int d\tau \sqrt{g} \left( g^{\tau\tau} \frac{\partial x}{\partial \tau} \cdot \frac{\partial x}{\partial \tau} - 1 \right), \tag{14}$$

is explicitly invariant under reparametrizations of the worldline. For instance, an infinitesimal diffeomorphism produced by a worldline vector $\varepsilon^{\tau 1}$ produces the following transformations:

$$\delta g_{\tau\tau} = 2 g_{\tau\tau} \partial_\tau \varepsilon^\tau + \varepsilon^\tau \partial_\tau g_{\tau\tau}, \quad \delta x^\mu = \varepsilon^\tau \partial_\tau x^\mu, \tag{15}$$

which leaves the action invariant up to a total derivative proportional to the Lagrangian

$$\delta S = \int d\tau \partial_\tau (\varepsilon^\tau L). \tag{16}$$

The Noether charge associated to these transformations, $G_L = p\, \delta x - \varepsilon^\tau L$, can be written as follows:

$$G_L = \frac{\sqrt{g}}{2}\left(g^{\tau\tau}\frac{\partial x}{\partial \tau}\cdot\frac{\partial x}{\partial \tau} + 1\right)\varepsilon. \tag{17}$$

*3.1. The Einbein Formulation: Lagrangian Symmetries.* Let us apply Castellani's procedure to the einbein formulation of the relativistic particle. Since there is only one component in the worldline metric, the action (14) is commonly given in terms of the einbein $e = \sqrt{g_{\tau\tau}}^{-2}$:

$$S = \frac{1}{2}\int d\tau\, e\left(e^{-2}\dot{x}^2 - 1\right). \tag{18}$$

In this formulation, the equations of motion give the following relations:

$$L_e \coloneqq \frac{1}{2}\left(e^{-2}\dot{x}^2 + 1\right) = 0, \quad L_x^{\mu} \coloneqq \frac{d}{dt}\left(e^{-1}\dot{x}\right) = 0. \tag{19}$$

It is easy to show that the Euler derivatives above satisfy the following Noether identity:

$$\tilde{\Lambda} \coloneqq \frac{d}{dt}L_e - e^{-1}\dot{x}_{\mu}L_x^{\mu} = 0. \tag{20}$$

Now, comparing this equation with (7), it is straightforward to read the associated functions $\rho$ as

$$\rho_e^1 = \delta(\tau), \quad \rho_x^0 = -e^{-1}\dot{x}. \tag{21}$$

Substituting the equations above in (6), we obtain the following transformations:

$$\delta e = \dot{\tilde{\varepsilon}}, \quad \delta x^{\mu} = \tilde{\varepsilon}e^{-1}\dot{x}^{\mu}, \tag{22}$$

which are symmetries of (18) as can be easily verified through direct calculation. However, we should note that (22) does not correspond to reparametrizations generated by a vector field on the worldline. Furthermore, the algebra of these transformations is abelian.

This unexpected result can be understood geometrically. To illustrate this, let us perform dimensional analysis on transformations (22). The variable $x$ maps space-time coordinates and $\tau$ stands for proper time, both possessing dimension 1. Additionally, $g_{\tau\tau}$ has dimension −2, indicating that the einbein $e$ has dimension −1. In order to restore symmetry through reparametrizations ($\tau \longrightarrow \tau + \varepsilon$), the gauge parameters must have dimension 1. However, our definition of $\tilde{\Lambda}$ assumes that the gauge parameter $\tilde{\varepsilon}$ has dimension 0, as evident from the comparison with (4).

We can amend this failure by introducing a new Noether identity:

$$\Lambda = e\tilde{\Lambda} = e\frac{d}{dt}L_e - \dot{x}.L_x = 0, \tag{23}$$

which clearly has dimension −1. In this scenario, following the standard procedure (7), we now obtain

$$\rho_e^1 = e, \quad \rho_e^0 = -\dot{e}, \quad \rho_x^0 = \dot{x}. \tag{24}$$

Substituting the functions above in (6), we obtain the expected transformations:

$$\delta e = e\dot{\varepsilon} + \varepsilon\dot{e}, \quad \delta x^{\mu} = \varepsilon\dot{x}^{\mu}, \tag{25}$$

which generate the standard algebra of diffeomorphism $\{\delta_{\eta}, \delta_{\varepsilon}\} = \delta_{[\eta,\varepsilon]}$.

*3.2. The Einbein Formulation: Hamiltonian Symmetries.* We now review the Hamiltonian formulation of the relativistic particle. We start by defining the conjugate momenta of $x$ and $e$, respectively:

$$p^{\mu} = e^{-1}\dot{x}^{\mu}, \quad \pi = 0, \tag{26}$$

which gives us our primary constraint $\Phi_1 = \pi$. It follows that the total Hamiltonian can be written as

$$H_T = \frac{e}{2}\left(p^2 + 1\right) + \dot{e}\pi. \tag{27}$$

Imposing consistency conditions on the constraint subspace leads to the secondary constraint $\Phi_2 = (1/2)(p^2 + 1)$. There are no tertiary constraints since $\{\Phi_2, H\} = 0$, and because of the relation $[\Phi_1, \Phi_2] = 0$, we can easily verify the closure of the constraint algebra. We can now apply Castellani's procedure to construct the generators of symmetries [9]. Firstly, we write an ansatz for the generator as a chain of linear combinations of first-class constraints as follows:

$$G(t) = \varepsilon_1(t)\pi + \frac{\varepsilon_2(t)}{2}\left(p^2 + 1\right). \tag{28}$$

Then, we identify the generator $G$ as a conserved charge by imposing Condition (13). Due to this requirement, we can write $\varepsilon_1$ in terms of $\varepsilon_2$. It follows that, the condition above implies that $\dot{\varepsilon}_2 = \varepsilon_1$, leading once again to the transformations:

$$\delta e = \dot{\varepsilon}, \quad \delta x^{\mu} = \varepsilon e^{-1}\dot{x}^{\mu}. \tag{29}$$

It is important to note from the discussion above that one drawback of the einbein formulation is the lack of covariance. One way to address this ambiguity is to explicitly use the action given by (14) right from the start of our Hamiltonian analysis. From the Hamiltonian perspective, this approach involves a transformation of the phase space variables:

$$(e, \pi) \longrightarrow (g_{\tau\tau}, p^{\tau\tau}), \tag{30}$$

and a redefinition of the canonical Hamiltonian as follows:

$$H = \frac{\sqrt{g_{\tau\tau}}}{2}\left(p^2 + 1\right). \tag{31}$$

In order for (30) to be a canonical transformation, two conditions must be satisfied [27]. Firstly, the Poincaré-Cartan 1-form,

$$\pi\,\delta e = p^{\tau\tau}\,\delta g_{\tau\tau}, \tag{32}$$

must remain unchanged. This can be achieved by setting the conjugate momentum of the metric as

$$p^{\tau\tau} = \frac{1}{2}\frac{\pi}{\sqrt{g_{\tau\tau}}}. \tag{33}$$

Secondly, Transformation (30) must preserve the total Hamiltonian, ensuring consistent dynamics. This requirement can easily be achieved since $\dot{g}_{\tau\tau}p^{\tau\tau} = \dot{e}\pi$. The total Hamiltonian can now be expressed as

$$H_T = \frac{\sqrt{g_{\tau\tau}}}{2}\left(p^2 + 1\right) + \dot{g}_{\tau\tau}p^{\tau\tau}. \tag{34}$$

The secondary constraint $\Phi^\tau = (1/2\sqrt{g_{\tau\tau}})(p^2 + 1)$ emerges after applying the consistency condition. The ansatz for the generator of symmetries naturally follows

$$G = \varepsilon_{\tau\tau}p^{\tau\tau} + \varepsilon_\tau\Phi^\tau = \varepsilon_{\tau\tau}p^{\tau\tau} + g_{\tau\tau}\varepsilon^\tau\Phi^\tau. \tag{35}$$

After applying Condition (13) to Generator (35), we arrive at the condition $\varepsilon_{\tau\tau} = 2g_{\tau\tau}\partial_\tau\varepsilon^\tau + \varepsilon^\tau\partial_\tau g_{\tau\tau}$, leading to the expected diffeomorphism transformations:

$$\delta g_{\tau\tau} = 2g_{\tau\tau}\partial_\tau\varepsilon^\tau + \varepsilon^\tau\partial_\tau g_{\tau\tau}, \quad \delta x^\mu = \varepsilon^\tau\partial_\tau x^\mu. \tag{36}$$

Taking into account this result, the einbein transformations (25) can be obtained easily since $\delta e = \delta\sqrt{g_{\tau\tau}}$.

The same transformations could also be obtained within the einbein formalism by redefining the primary constraint as $\Phi_1 = e\pi$. As a consequence, the canonical Hamiltonian emerges as the secondary constraint, $\Phi_2 = (e/2)(p^2 + 1)$. The generator of symmetries is now given by

$$G(t) = \varepsilon^{(1)}(t)e\,\pi + \frac{\varepsilon^{(2)}(t)}{2}e\left(p^2 + 1\right). \tag{37}$$

Imposing (13) to the generator above, one gets the relation $e\varepsilon^{(1)} = e\dot{\varepsilon}^{(2)} + \varepsilon^{(2)}\dot{e}$, which is consistent with the Lagrangian transformations (25).

Why is it necessary to redefine $\phi_1$ in order to obtain the correct reparametrization transformations? Due to the equivalence between Lagrangian and Hamiltonian formulations, it is essential that the pullback of the Hamiltonian generator of symmetries $G$ matches the Lagrangian Noether charge (17). It is clear that $FL^*G(t) = (\varepsilon(t)/2)(e^{-1}\dot{x}^2 + e)$, as required.

Based on the preceding discussion, it is evident that the structure of gauge transformations has nothing to do with the canonicity of the phase space variables. Instead, it relies on the nature of the local gauge parameters. From the Hamiltonian point of view, the choice of primary constraints may disrupt covariance, thereby influencing the nature of gauge parameters.

## 4. Hamiltonian Formulation of the Polyakov String

The action of the Polyakov string is given by

$$S = \frac{1}{2}\int_\Sigma d^2\sigma\sqrt{-g}\,g^{\alpha\beta}\partial_\alpha x^\mu\partial_\beta x^\nu\eta_{\mu\nu}, \tag{38}$$

where $\mu, \nu = 0, 1, \cdots, D-1$ are target space coordinates and $\alpha, \beta = 0, 1$ are worldsheet coordinates. The Hamiltonian analysis starts by defining the canonical momenta:

$$\mathscr{P}_\mu = \frac{\partial L}{\partial(\partial_0 x^\mu)} = \sqrt{-g}g^{o\alpha}\partial_\alpha x_\mu, \tag{39}$$

and

$$\pi^{\alpha\beta} = \frac{\partial L}{\partial\left(\partial_0 g_{\alpha\beta}\right)} = 0, \tag{40}$$

where we can identify three independent primary constraints:

$$\pi^{00} = 0, \quad \pi^{01} = 0, \quad \pi^{11} = 0. \tag{41}$$

It is convenient to rewrite the velocities $\partial_0 x^\mu$ in terms of $\mathscr{P}_\mu$:

$$\partial_0 x^\mu = \frac{1}{g^{00}}\left(\frac{\mathscr{P}^\mu}{\sqrt{-g}} - g^{01}\partial_1 x^\mu\right). \tag{42}$$

Then, the canonical Hamiltonian density is given by

$$\mathscr{H}_c = -\frac{1}{\sqrt{-g}g^{00}}H - \frac{g^{01}}{g^{00}}T, \tag{43}$$

where

$$H = \frac{1}{2}\left(\mathscr{P}^2 + {x'}^2\right), \quad T = \mathscr{P}^\mu x'_\mu. \tag{44}$$

We define the basic Poisson brackets as

$$\left\{x^\mu(\sigma), \mathscr{P}^\nu\left(\sigma'\right)\right\} = \eta^{\mu\nu}\delta\left(\sigma - \sigma'\right), \tag{45}$$

$$\left\{g_{\rho\theta}(\sigma), \pi^{\alpha\beta}\left(\sigma'\right)\right\} = \frac{1}{2}\left(\delta^\alpha_\rho\delta^\beta_\theta + \delta^\alpha_\theta\delta^\beta_\rho\right)\delta\left(\sigma - \sigma'\right). \tag{46}$$

All other fundamental Poisson brackets between the variables of the metric vanish.

Since we will need to compute Poisson brackets between the Hamiltonian and $\pi^{\alpha\beta}$, it is convenient to rewrite (43) in terms of the covariant components of the metric:

$$\mathscr{H}_c = -\frac{\sqrt{-g}}{g_{11}}H + \frac{g_{01}}{g_{11}}T, \tag{47}$$

where we have defined $g^{\alpha\beta}$ as the inverse of $g_{\alpha\beta}$, so that

$$\begin{pmatrix} g^{00} & g^{01} \\ g^{10} & g^{11} \end{pmatrix} = \frac{1}{g}\begin{pmatrix} g_{11} & -g_{01} \\ -g_{10} & g_{00} \end{pmatrix}. \tag{48}$$

The total Hamiltonian, which is given by

$$H_T = \int d\sigma\left(\mathscr{H}_c + \lambda_{\alpha\beta}\pi^{\alpha\beta}\right), \tag{49}$$

describes the dynamics of the theory. This leads to the following equations of motion:

$$\begin{aligned} \dot{x}^\mu &= \{x^\mu, H_T\} = \frac{g_{01}}{g_{11}}x'^\mu - \frac{\sqrt{-g}}{g_{11}}\mathscr{P}^\mu, \\ \dot{\mathscr{P}}^\mu &= \{\mathscr{P}^\mu, H_T\} = -\partial_1\left(\frac{\sqrt{-g}}{g_{11}}x'^\mu + \frac{g_{01}}{g_{11}}\mathscr{P}^\mu\right), \\ \dot{g}_{\alpha\beta} &= \left\{g_{\alpha\beta}, H_T\right\} = \lambda_{\alpha\beta}. \end{aligned} \tag{50}$$

## 5. Closure of the Dirac Procedure

After imposing the stability condition on the primary constraints $\dot{\pi}^{\alpha\beta} = \{\pi^{\alpha\beta}, H_c\}$, secondary constraints emerge. Considering $\{\sqrt{-g}, \pi^{\alpha\beta}\} = (1/2)\sqrt{-g}\, g^{\alpha\beta}$, the three secondary constraints can be written as

$$\begin{aligned} \dot{\pi}^{00} &= \frac{-1}{2\sqrt{-g}}H, \\ \dot{\pi}^{10} &= \frac{g_{01}}{2\sqrt{-g}g_{11}}H - \frac{1}{2g_{11}}T. \\ \dot{\pi}^{11} &= \frac{1}{\sqrt{-g}}\left(\frac{g}{(g_{11})^2} - \frac{g_{00}}{2g_{11}}\right)H + \frac{g_{01}}{g_{11}^2}T. \end{aligned} \tag{51}$$

Not all these constraints are linearly independent; in fact, they are related by the following identity:

$$g_{\alpha\beta}\dot{\pi}^{\alpha\beta} = g_{00}\dot{\pi}^{00} + 2g_{01}\dot{\pi}^{01} + g_{11}\dot{\pi}^{11} = 0. \tag{52}$$

Upon closer examination of (51), it becomes clear that $T$ and $H$ are the only independent functions in the space of secondary constraints and span the following closed algebra:

$$\begin{aligned} \left\{H(\sigma), H\left(\sigma'\right)\right\} &= \left(T(\sigma) + T\left(\sigma'\right)\right)\partial_1\delta\left(\sigma - \sigma'\right), \\ \left\{H(\sigma), T\left(\sigma'\right)\right\} &= \left(H(\sigma) + H\left(\sigma'\right)\right)\partial_1\delta\left(\sigma - \sigma'\right), \\ \left\{T(\sigma), T\left(\sigma'\right)\right\} &= \left(T(\sigma) + T\left(\sigma'\right)\right)\partial_1\delta\left(\sigma - \sigma'\right). \end{aligned} \tag{53}$$

We can notice that the secondary constraints are also first class. Moreover, no tertiary constraints arise since

$$\begin{aligned} \{H(\sigma), H_c\} &= -2T\,\partial_1\left(\frac{\sqrt{-g}}{g_{11}}\right) - \left(\frac{\sqrt{-g}}{g_{11}}\right)\partial_1 T + 2H\,\partial_1\left(\frac{g_{01}}{g_{11}}\right) + \left(\frac{g_{01}}{g_{11}}\right)\partial_1 H, \\ \{T(\sigma), H_c\} &= -2H\,\partial_1\left(\frac{\sqrt{-g}}{g_{11}}\right) - \left(\frac{\sqrt{-g}}{g_{11}}\right)\partial_1 H + 2T\,\partial_1\left(\frac{g_{01}}{g_{11}}\right) + \left(\frac{g_{01}}{g_{11}}\right)\partial_1 T. \end{aligned} \tag{54}$$

## 6. The Generator

Our model features five independent first-class constraints. While it may initially seem natural to choose the three primary constraints along with the secondary constraints $T$ and $H$, this approach does not adhere to the general procedural rules, breaking the covariance of the constraints, and potentially leading to noncovariant transformations. In our analysis, we propose an alternative approach by directly working with the three primary constraints (41) and the three secondary constraints (51) which transforms covariantly as tensor densities.

Based on the discussion above, we propose the following structure for the generator of symmetries:

$$G = \int d\sigma\left[\varepsilon_{\alpha\beta}\pi^{\alpha\beta} + \dot{\pi}^{0\alpha}\varepsilon_\alpha\right]. \tag{55}$$

It is important to remark that this generator is unique up to a term proportional to the Weyl identity (52), so that we can always rewrite $\dot{\pi}^{11}$ in terms of $\dot{\pi}^{00}$ and $\dot{\pi}^{01}$ in the generator above.

In order to compare with the standard form of diffeomorphism transformations, it is convenient to express (55) in terms of contravariant worldsheet vectors. Defining $\phi_\alpha = g_{\alpha\beta}\dot{\pi}^{0\beta}$, it follows that

$$G = \int d\sigma\left[\varepsilon_{\alpha\beta}\pi^{\alpha\beta} + \varepsilon^\alpha\phi_\alpha\right]. \tag{56}$$

Using Equations (51), we can explicitly rewrite the components of $\phi_\alpha$ as

$$\begin{aligned} \phi_0 &= \frac{\sqrt{-g}}{g_{11}}H - \frac{g_{01}}{g_{11}}T = -\mathscr{H}_c, \\ \phi_1 &= -T, \end{aligned} \tag{57}$$

which satisfy the following Poisson brackets:

$$\begin{aligned} \{H_c, \phi_0\} = \int d\sigma' \bigg[ &\left(\partial_1\left(\frac{g_{00}}{g_{11}}\right)\delta\left(\sigma - \sigma'\right) - \left(\frac{g_{00}}{g_{11}}\right)\partial_1'\delta\left(\sigma - \sigma'\right)\right)\phi_1 \\ &- 2\left(\partial_1\left(\frac{g_{01}}{g_{11}}\right)\delta\left(\sigma - \sigma'\right) - \left(\frac{g_{01}}{g_{11}}\right)\partial_1'\delta\left(\sigma - \sigma'\right)\right)\phi_0\bigg], \end{aligned} \tag{58}$$

and

$$\{H_c,\phi_1\} = \int d\sigma'\left[\left(-\frac{g_{11}}{\sqrt{-g}}\partial_1\left(\frac{\sqrt{-g}}{g_{11}}\right)\delta\left(\sigma-\sigma'\right)+\partial_1'\delta\left(\sigma-\sigma'\right)\right)\phi_0 + \left(\frac{g_{01}}{\sqrt{-g}}\partial_1\left(\frac{\sqrt{-g}}{g_{11}}\right)-\partial_1\left(\frac{g_{01}}{g_{11}}\right)\right)\delta\left(\sigma-\sigma'\right)\phi_1\right]. \tag{59}$$

To express $\varepsilon_{\alpha\beta}$ in terms of the worldsheet vector $\varepsilon^\alpha$, we need to ensure that Generator (56) satisfies the master Equation (13). The resulting condition can be written as

$$0 = \partial_0\varepsilon^\alpha\phi_\alpha + \dot{\pi}^{\alpha\beta}\varepsilon_{\alpha\beta} + \varepsilon^\alpha\{\phi_\alpha, H_c\} + \varepsilon^\gamma\partial_0 g_{\alpha\beta}\left\{\phi_\gamma,\pi^{\alpha\beta}\right\}. \tag{60}$$

Let us compute each of the Poisson brackets that appear in (60) separately. The first contribution can be easily computed by taking into account (58) and (59), that is,

$$\begin{aligned}\varepsilon^\alpha\{\phi_\alpha, H_c\} = \int d\sigma\Bigg[&\partial_1\varepsilon^0\left(\frac{g_{00}}{g_{11}}\phi_1 - 2\frac{g_{01}}{g_{11}}\phi_0\right)\\ &-\partial_1\varepsilon^1\phi_0 ++\varepsilon^1\left(\frac{g_{01}}{\sqrt{-g}}\partial_1\left(\frac{\sqrt{-g}}{g_{11}}\right)\right)\phi_0\\ &-\left(\partial_1\left(\frac{g_{01}}{g_{11}}\right)-\frac{g_{11}}{\sqrt{-g}}\partial_1\left(\frac{\sqrt{-g}}{g_{11}}\right)\right)\phi_1\Bigg].\end{aligned} \tag{61}$$

With the help of (52), we can write (61) in terms of $\dot{\pi}^{\alpha\beta}$ as

$$\begin{aligned}\varepsilon^\alpha\{\phi_\alpha, H_c\} = 2\int d\sigma\big[&\partial_1\varepsilon^1\left(g_{01}\dot{\pi}^{01}+g_{11}\dot{\pi}^{11}\right)+\partial_1\varepsilon^0\left(g_{00}\dot{\pi}^{01}+g_{01}\dot{\pi}^{01}\right)\\ &++\varepsilon^1\left(\partial_1 g_{00}\dot{\pi}^{00}+2\partial_1 g_{01}\dot{\pi}^{01}+\partial_1 g_{11}\dot{\pi}^{11}\right)\big].\end{aligned} \tag{62}$$

Taking into account that $\{\phi_1,\pi^{\alpha\beta}\}=0$, the last contribution in (60) can be simplified as follows:

$$\varepsilon^0\partial_0 g_{\alpha\beta}\left\{\phi_0,\pi^{\alpha\beta}\right\} = \dot{\pi}^{\alpha\beta}\left(\varepsilon^0\partial_0 g_{\alpha\beta}\right). \tag{63}$$

Now, we can join all the contributions to (60), obtaining

$$0 = \dot{\pi}^{\alpha\beta}\left(\varepsilon_{\alpha\beta}+2\partial_\alpha\varepsilon^\gamma g_{\beta\gamma}+\varepsilon^\gamma\partial_\gamma g_{\alpha\beta}\right). \tag{64}$$

Because of identity (52), the term between the brackets must be equal to zero up to a term proportional to $g_{\alpha\beta}$. As you may suspect, this is the source of Weyl symmetry. Then, it follows that

$$\varepsilon_{\alpha\beta} = g_{\gamma(\beta}\partial_{\alpha)}\varepsilon^\gamma + \varepsilon^\gamma\partial_\gamma g_{\alpha\beta} + \Omega\, g_{\alpha\beta}. \tag{65}$$

Finally, we can write the Hamiltonian generator of symmetries as

$$G_\varepsilon(t) = \int d\sigma\left[\left(g_{\gamma(\beta}\partial_{\alpha)}\varepsilon^\gamma + \varepsilon^\gamma\partial_\gamma g_{\alpha\beta} + \Omega\, g_{\alpha\beta}\right)\pi^{\alpha\beta} + \varepsilon^\alpha\phi_\alpha\right], \tag{66}$$

where $\Omega$ is an arbitrary function of the worldsheet variables. With this generator at hand, we can obtain the transformation of the metric tensor:

$$\delta_\varepsilon g_{\alpha\beta} = g_{\gamma(\beta}\partial_{\alpha)}\varepsilon^\gamma + \varepsilon^\gamma\partial_\gamma g_{\alpha\beta} + \Omega\, g_{\alpha\beta}, \tag{67}$$

and the transformation of the matter field

$$\delta_\varepsilon x^\mu = \varepsilon^0\left(\frac{-\sqrt{-g}P^\mu + g_{01}\partial_\sigma x^\mu}{g_{11}}\right) + \varepsilon^1\partial_\sigma x^\mu = \varepsilon^\alpha\partial_\alpha x^\mu. \tag{68}$$

These results show that the symmetries within the Hamiltonian formulation produce identical covariant transformations as those found in the Lagrangian description. Moreover, the generators (66) satisfy the following nontrivial Poisson bracket algebra:

$$\left\{G_\varepsilon, G_\eta\right\} = G_{[\varepsilon,\eta]}, \tag{69}$$

which serves as a representation of the local symmetry algebra. It takes quite a long calculation to check this result. Let us check this assertion by explicitly writing each contribution.

First of all, considering transformations of the kind (67), the generators of the metric transformations reproduce the expected subalgebra:

$$\int d\sigma\int d\sigma'\left\{\delta_\varepsilon g_{\alpha\beta}\pi^{\alpha\beta}(\sigma), \delta_\eta g_{\rho\lambda}\pi^{\rho\lambda}\left(\sigma'\right)\right\} = \int d\sigma\delta_{[\varepsilon,\eta]}g_{\alpha\beta}\pi^{\alpha\beta}(\sigma). \tag{70}$$

On the other hand, there is a mixed contribution:

$$\begin{aligned}&\int d\sigma\int d\sigma'\left[\left\{\delta_\varepsilon g_{\alpha\beta}\pi^{\alpha\beta}(\sigma), \eta^\alpha\phi_\alpha\left(\sigma'\right)\right\} + \left\{\varepsilon^\alpha\phi_\alpha(\sigma), \delta_\eta g_{\alpha\beta}\pi^{\alpha\beta}\left(\sigma'\right)\right\}\right]\\ &\quad= \int d\sigma\Big[\left(\varepsilon^0\partial_0\eta^\alpha - \eta^0\partial_0\varepsilon^\alpha\right)\phi_\alpha + \left(\varepsilon^0\eta^1 - \eta^0\varepsilon^1\right)\partial_1 g_{\alpha\beta}\dot{\pi}^{\alpha\beta}\\ &\quad+2\left(\varepsilon^0\partial_1\eta^\alpha - \eta^0\partial_1\varepsilon^\alpha\right)g_{\alpha\lambda}\dot{\pi}^{\alpha 1}\Big],\end{aligned} \tag{71}$$

and the algebra of the matter field transformations

$$\begin{aligned}&\int d\sigma\int d\sigma'\left\{\varepsilon^\alpha\phi_\alpha(\sigma), \eta^\alpha\phi_\alpha\left(\sigma'\right)\right\}\\ &\quad= \int d\sigma\Big[\left(\varepsilon^1\partial_1\eta^\alpha - \eta^1\partial_1\varepsilon^\alpha\right)\phi_\alpha - \left(\varepsilon^0\eta^1 - \eta^0\varepsilon^1\right)\partial_1 g_{\alpha\beta}\dot{\pi}^{\alpha\beta}\\ &\quad-2\left(\varepsilon^0\partial_1\eta^\alpha - \eta^0\partial_1\varepsilon^\alpha\right)g_{\alpha\lambda}\dot{\pi}^{\alpha 1}\Big],\end{aligned} \tag{72}$$

which together reproduce the algebra (69), thereby confirming the consistency of the Hamiltonian gauge transformations.

## 7. Hamiltonian Analysis in the Lapse and Shift Variables

It is time to make contact with the standard discussion on the Hamiltonian theory of bosonic string theory which is usually made in terms of lapse and shift variables [14, 16]. A widely acknowledged result is that the symmetry transformations derived from this formulation do not reproduce the usual diffeomorphism transformations. Consequently, in order to recover the usual structures of diffeomorphisms, a field-dependent redefinition of the gauge parameters is needed. In this section, we will explore the causes behind this failure. We will show that this result has nothing to do with the canonicity of the variables but arises because the conjugate momenta associated with the shift and lapse variables are not worldsheet tensors. Nevertheless, the transformations derived using Castellani's procedure remain symmetries of the theory, even though a real worldsheet vector field does not induce them.

*7.1. Canonicity of the Transformation.* It is possible to simplify the canonical Hamiltonian density (47) by parametrizing the worldsheet metric defining lapse and shift functions as follows:

$$N = -\frac{\sqrt{-g}}{g_{11}}, \quad N_1 = \frac{g_{01}}{g_{11}}, \quad \Omega = g_{11}. \tag{73}$$

In this parametrization, the metric is written as

$$g_{\alpha\beta} = \Omega \begin{pmatrix} N_1^2 - N^2 & N_1 \\ N_1 & 1 \end{pmatrix}. \tag{74}$$

Then, the Hamiltonian can be written in the following way:

$$\mathscr{H}_T = NH + N_1 T. \tag{75}$$

Given the transformations (73), we can find the corresponding transformations of the momenta that preserve the Poisson brackets. This can be done by imposing the invariance of the Poincaré-Cartan 1-form:

$$\pi^{\alpha\beta}\delta g_{\alpha\beta} = \Pi\delta N + \Pi_1 \delta N_1 + \Pi_{11}\delta\Omega. \tag{76}$$

Taking the variational derivative with respect to $g_{\alpha\beta}$, we obtain

$$\begin{aligned} \pi^{00} &= \frac{1}{2\sqrt{-g}}\Pi, \\ \pi^{01} &= \frac{g_{01}}{2g_{11}\sqrt{-g}}\Pi + \frac{1}{2g_{11}}\Pi_1, \\ \pi^{11} &= \frac{\left(g_{01}^2 - (1/2)g_{00}g_{11}\right)}{g_{11}^2\sqrt{-g}}\Pi - \frac{g_{01}}{g_{11}^2}\Pi_1 + \Pi_{11}. \end{aligned} \tag{77}$$

Inverting these relations, the momenta $\Pi$ can be written as

$$\begin{aligned} \Pi &= 2\sqrt{-g}\pi^{00}, \\ \Pi_1 &= 2g_{11}\pi^{01} + 2g_{01}\pi^{00}, \\ \Pi_{11} &= \frac{1}{g_{11}} g_{\alpha\beta}\pi^{\alpha\beta}. \end{aligned} \tag{78}$$

In order to preserve the dynamics of the theory, the total Hamiltonian, which includes the primary constraints, must preserve its structure. This is indeed the case because

$$\dot{g}_{\alpha\beta}\pi^{\alpha\beta} = \Pi\dot{N} + \Pi_1\dot{N}_1 + \Pi_{11}\dot{\Omega}. \tag{79}$$

From this result, we can conclude that the transformations (73) are canonical.

*7.2. Closure of the Dirac Procedure.* From our previous redefinition of the conjugate momenta (78), it is clear that the first-class constraints remain as

$$\begin{aligned} \Pi &\approx 0, \\ \Pi_1 &\approx 0, \\ \Pi_{11} &\approx 0. \end{aligned} \tag{80}$$

Applying the consistency algorithm to the primary constraints results in the following secondary constraints:

$$\begin{aligned} \{H_{can}, \Pi\} &= H, \\ \{H_{can}, \Pi_1\} &= T, \\ \{H_{can}, \Pi_{11}\} &= 0. \end{aligned} \tag{81}$$

No tertiary constraints arise due to (54), confirming the closure of the Dirac procedure.

*7.3. The Generator.* Following Castellani's prescription, we can construct the Hamiltonian generator as a linear combination of the first-class constraints:

$$G_\nu = \int d\sigma \left(\lambda\Pi + \lambda_1\Pi_1 + \lambda_{11}\Pi_{11} + \nu^\perp H + \nu^1 T\right). \tag{82}$$

The main difference between this generator and the one utilized in (66) is that none of the first-class constraints in (82) are worldsheet tensors. Consequently, there is no guarantee that the related gauge parameters are indeed vector fields.

Imposing Castellani's Condition (13) on Generator (82), we can express the variables $\lambda$ in terms of the gauge parameters $\nu^\perp$ and $\nu^1$. This condition can be solved by considering the commutation Relations (53), (54), and (81), we get

$$\begin{aligned} \lambda_1 &= \dot{\nu}^1 + \nu^1\partial_1 N_1 - N_1\partial_1\nu^1 + \nu^\perp\partial_1 N - N\partial_1\nu^\perp, \\ \lambda &= \dot{\nu}^\perp + \nu^1\partial_1 N - N\partial_1\nu^1 + \nu^\perp\partial_1 N_1 - N_1\partial_1\nu^\perp. \end{aligned} \tag{83}$$

As a consequence, we can derive the transformations for the lapse and shift function as

$$\delta N = \{N, G\} = \lambda, \quad \delta N_1 = \{N_1, G\} = \lambda_1, \tag{84}$$

and when Expression (82) acts on the matter fields, we observe the following variations:

$$\delta x^{\mu} = \{x^{\mu}, G\} = \nu^{\perp} \mathscr{P}^{\mu} + \nu^1 \partial_1 x^{\mu}. \tag{85}$$

It is straightforward to show that the generator performs the following algebra:

$$\{G_{\nu}, G_{w}\} = G_{[\nu,w]} \tag{86}$$

such that,

$$[\nu, w]^1 = \left(\nu^1 \partial_1 w^1 + \nu^{\perp} \partial_1 w^{\perp}\right) - (\nu \leftrightarrow w), \tag{87}$$

$$[\nu, w]^{\perp} = \left(\nu^1 \partial_1 w^{\perp} + \nu^{\perp} \partial_1 w^1\right) - (\nu \leftrightarrow w). \tag{88}$$

The above relations demonstrate that the algebra of $G_{\nu}$ closes off-shell. However, when pulled back to configuration space, Transformation (85) does not reproduce the standard diffeomorphism form. For instance,

$$\delta x^{\mu} = \left(\nu^1 - \nu^{\perp} \frac{g_{01}}{\sqrt{-g}}\right) \partial_1 x^{\mu} - \nu^1 \frac{g_{11}}{\sqrt{-g}} \partial_0 x^{\mu}. \tag{89}$$

This suggests that the canonical representation of diffeomorphisms must necessarily involve field-dependent parameters when expressed in ADM variables. The correspondence with spacetime diffeomorphisms can nevertheless be restored by introducing the lapse–shift decomposition:

$$\nu^{\perp} = \varepsilon^0 N, \quad \nu^1 = \varepsilon^1 + N_1 \varepsilon^0, \tag{90}$$

which represents a field-dependent redefinition of the gauge parameters. This feature, first noted by Kuchař [15] and later made explicit in the analyses of Brink and Henneaux [14] and Pons et al. [16], clarifies how the ADM parametrization realizes diffeomorphisms only after identifying the phase space generators with the spacetime vector fields through such redefinitions.

## 8. Concluding Remarks

By applying Castellani's procedure, we have shown how to construct the Hamiltonian generator of symmetries for bosonic string theory. The resulting transformations are the same as those of the Lagrangian formulation[3]. The worldsheet metric does not possess dynamics and works as an auxiliary field for the classical theory. Consequently, the elements of its conjugate momenta $\pi^{\alpha\beta}$ represent the primary first-class constraints of the model, which manifestly transform as worldsheet tensor densities, as well as the secondary constraints $\dot{\pi}^{\alpha\beta}$, that arise as a consequence of the consistency algorithm.

In the Hamiltonian analysis of bosonic string theory, a key feature is that the Hamiltonian time evolution vector field $X_H$ (9) has a kernel on the pfcc space given by elements proportional to the trace of the conjugate momenta $\pi^{\alpha\beta}$. This result offers two approaches for analyzing the secondary constraints. One option is simplifying the constraint space by considering only two independent secondary constraints, $T$ and $H$. However, this choice comes at the expense of losing covariance on the constraint space. The alternative approach preserves covariance by maintaining the three components of $\dot{\pi}^{\alpha\beta}$ as secondary constraints, related by the Noether identity (52), which give rise to Weyl symmetry. Our work demonstrates that this is the appropriate criterion for deriving diffeomorphism transformations for bosonic string theory.

There have been recent attempts to address related questions by focusing directly on the algebra of symmetries. For instance, Banerjee et al. [28] studied the nonrelativistic bosonic string in flat spacetime, while Nandi et al. [29] extended the analysis to Newton–Cartan backgrounds. Although these works represent valuable contributions, they still rely on constraint structures that obscure covariance in a manner reminiscent of the ADM-type parametrizations discussed above. In contrast, our approach contrasts sharply with such treatments by reconstructing the covariant form of the symmetries directly from Castellani's algorithm when applied to genuinely covariant chains of constraints.

In a broader sense, our research has shown that certain models, even when related by canonical transformations, do not reproduce the correct Hamiltonian gauge transformations. This is particularly evident in the case of the relativistic particle and the bosonic string. Understanding the conditions under which two canonically related theories exhibit the same Hamiltonian gauge transformations could have significant implications. Castellani's procedure fails to guarantee this match, opening the door to further discussions and deeper analysis in the field. This observation resonates with the broader issue of whether canonical equivalence necessarily implies gauge equivalence, a subtlety that deserves further attention.

Finally, let us briefly comment on possible implications for quantization. Once a gauge is fixed, the Hamiltonian constraints considered here reduce to the familiar Virasoro algebra (53), which is the usual starting point for the BRST quantization of the bosonic string. In this respect, our construction provides a complementary view of how the Virasoro structure emerges from the more general diffeomorphism algebra at the Hamiltonian level. Moreover, since our analysis has been entirely Hamiltonian, a natural extension would be to explore its continuation within the Batalin–Fradkin–Vilkovisky (BFV) formalism [30, 31], which may provide a bridge to the standard Lagrangian treatments, while also clarifying whether any modifications arise in connection with approaches that incorporate ghosts for the moduli space [32]. This perspective may help to sharpen the Hamiltonian foundations of BRST quantization in string theory.

## Data Availability Statement

Data sharing is not applicable to this article as no datasets were generated or analyzed during the current study.

## Disclosure

An earlier version of this work has been published as an arXiv preprint: arXiv:2406.14019v2 [33].

## Conflicts of Interest

The authors declare no conflicts of interest.

## Funding

No funding was received for this manuscript.

## Acknowledgments

We would like to express our deepest gratitude to professor Victor Rivelles, for his invaluable guidance and inspiration. His memory will always be cherished.

## Endnotes

[1]It may seem unnecessary to explicitly indicate the $\tau$ component for a one-dimensional vector field. However, it will help us remember that $\varepsilon^\tau$ is a proper vector field on the world line.

[2]Given the transformation (15) for the metric, it is easy to show that the einbein must transform as follows: $\delta e = e\dot{\varepsilon} + \varepsilon\dot{e}$.

[3]For a detailed explanation of how the symmetries of the Polyakov string are derived using the Lagrangian formalism, we refer the reader to [6], in which the authors have presented a detailed program to examine the gauge symmetries of a theory with a singular Lagrangian.